\documentclass[11pt]{amsart}
\usepackage{amssymb}
\usepackage{amsmath,amsthm,amsfonts}  
\usepackage[applemac]{inputenc} 
 \usepackage{graphicx}
\usepackage{amsthm}
\usepackage{bbm}
\usepackage{pdfsync}
\usepackage{fancyhdr}
\usepackage[T1]{fontenc}

\usepackage{color}
\usepackage{xcolor}
\usepackage[breaklinks,colorlinks,backref]{hyperref}
\hypersetup{
    colorlinks, 
    linktoc=all, 
    linkcolor=red,  
  citecolor=hyptxt,
  urlcolor=blue}
  \hypersetup{
  citebordercolor=,
  filebordercolor=green,
  linkbordercolor=blue
}
\definecolor{hyptxt}{rgb}{0.7, 0.4, 0.9}
\usepackage{bibtopic}

\newtheorem{prop}{Proposition}[section]

\newtheorem{form}{Formulae}[section]
\newenvironment{remark}[1][Remark]{\begin{trivlist}
\item[\hskip \labelsep {\bfseries #1}]}{\end{trivlist}}

\newcommand{\beprop}{\begin{prop}}
\newcommand{\enprop}{\end{prop}}
\newcommand{\befor}{\begin{form}}
\newcommand{\enfor}{\end{form}}
\newcommand{\bprf}{\begin{proof}} 
\newcommand{\eprf}{\end{proof}}

\definecolor{hervecolor}{rgb}{0.8,0,0.7}
     
\newcommand{\ket}[1]{|\kern.3ex#1\kern.3ex\rangle}
\newcommand{\bra}[1]{\langle\kern.3ex #1 \kern.3ex|}
\newcommand{\scalar}[2]{\langle\kern.3ex #1 \kern.3ex|\kern.3ex#2\kern.3ex\rangle}

\newcommand{\ii}{\mathsf{i}}

\def\R{\mathbb{R}}

\def\Z {\mathbb{Z}}

\def\lg{\langle }
\def\rg{\rangle }

\def\vap{\varpi}

\def\ud{\mathrm{d}}

\def\sfM{\mathsf{M}}
\def\sfP{\mathsf{P}}
\def\sfMv{\mathsf{M}^{\vap}}

\numberwithin{equation}{section}

\title{Integral Quantization for the Discrete Cylinder}
\author{Jean Pierre Gazeau and Romain Murenzi}
\address{Universit\'e Paris Cit\'e, CNRS, Astroparticule et Cosmologie, F-75013 Paris, France}\email{gazeau@apc.in2p3.fr}
\address{The World Academy of Sciences, TWAS/ICTP\\
Via Costiera 1, Trieste 34151, Italy
}
\email{rmurenzi@twas.org}

\date{\today}        

\begin{document}

\begin{abstract}
Covariant integral quantizations  are  based on the resolution of the identity by continuous or discrete families of normalised positive operator valued measures (POVM), which have appealing probabilistic content and  which transform in a covariant way. One of their advantages is to allow to  circumvent problems due to the presence of singularities in the classical models. In this paper we implement covariant  integral quantizations for  systems whose phase space is  $\Z\times\,\mathbb{S}^1$, i.e., for systems moving on the circle.  The symmetry  group of this phase space is the discrete \& compact  version of the Weyl-Heisenberg group, namely  the central extension of the abelian group $\Z\times\,\mathrm{SO}(2)$. In this regard, the phase space is viewed as the right coset of the group with its center. The non-trivial unitary irreducible representation of this group, as acting on $L^2(\mathbb{S}^1)$,  is square integrable on the phase space. We show how to derive  corresponding covariant  integral quantizations from (weight) functions on the phase space {and resulting resolution of the identity}. {As particular cases of the latter} we recover  quantizations with de Bi\`evre-del Olmo-Gonzales and Kowalski-Rembielevski-Papaloucas coherent states on the circle. Another straightforward outcome of our approach is  the Mukunda Wigner transform.  We also look at the specific cases of coherent states built from shifted gaussians, Von Mises,  Poisson, and Fej\'er kernels. Applications to stellar representations are in progress.
\end{abstract}

\maketitle



\tableofcontents

\section {Introduction}
\label{intro}
Quantum physics for systems whose configuration space is the circle $\mathbb{S}^1$  has been considered by many authors over the last six decades \cite{Carruthers,leblond,mukunda1,Floreanini,hall,Scardicchio,Gour,Zhang,Kowalski1,debievre,aremua,Rigasa,chadzitaskos,Przanowski,kastrup,Fresneda,Kowalski2,Mista,perel86,Gonzales}. {These studies include the work by Carruthers (1968) \cite{Carruthers}, L\'evy-Leblond (1976) \cite{leblond}, Kowalski \textit{et al} (1996 and 2021) \cite{Kowalski1,Kowalski2}, De Bi\`evre \cite{debievre}, Gazeau \textit{et al} \cite{aremua,Fresneda}, until the most recent one by Mista \textit{et al} \cite{Mista}}. The  phase space associated with  this configuration space is the cylinder, continuous or discrete. These works included questions such as, for a given quantization scheme of the phase space, 
\begin{itemize}
  \item[(i)] what is the appropriate quantum position operator? 
  \item[(ii)] what is the appropriate  quantum momentum operator? 
  \item[(iii)] what are the  states (coherent states) that saturate some uncertainty relations?
  \item[(iv)] do these operators explicitly display the topology of the circle or the fact that the circle is a curved manifold or a group?
\end{itemize} 
The later question raises interest on the extension of these investigations to higher dimensional configuration spaces, such as the group of rotation SO(3) in three dimensions. {Note that this group can be} viewed as  the configuration space for the motion of a rigid body about one of its points.

{Covariant integral quantizations \cite{bergaz14}  linearly transform functions (``classical observables'') on phase spaces (in a wider sense) into operators ``quantum observables'') on some Hilbert spaces of ``quantum states''. They  are  based on the resolution of the identity by continuous or discrete families of normalised positive operator valued measures (POVM) which transform in a covariant way  under some symmetry group actions. In the simplest cases these symmetries are described by the   Weyl-Heisenberg group (projective representations of translations in 2d dimensions), or by the affine groups (translations of a subset of variables combined with dilations of the remnant subset of bounded below variables). These quantization methods whose the origin can be traced back to Klauder, Berezin, Toepiltz, are relatively easy to manipulate when compared with geometric or deformation or other quantizations. Beyond their appealing probabilistic content, they  allow to  circumvent  ordering problems, or those due to the presence of singularities in the classical models.}
 
In this work, we implement the covariant integral quantizations for  systems whose phase space {or ``discrete cylinder''} is $\Z\times\,\mathbb{S}^1$, i.e., for systems moving on the circle and with {integral momentum}.  
The symmetry  group of this phase space is the discrete \& compact  version  of the Weyl-Heisenberg group H$_1$. More precisely, it is a  central extension of the abelian group $\Z\times\,\mathrm{SO}(2)$, homeomorphic to $\R\times\Z\times\,\mathrm{SO}(2)$, and here denoted by $\mathrm{H}^{\mathrm{dc}}_1$.   The phase space can be viewed as the {right coset $\Gamma:=\R\backslash\mathrm{H}^{\mathrm{dc}}_1$}.

In Section \ref{survey}  {we briefly review the  covariant integral quantization procedure. We particularly emphasize the role of the {resolution} of the identity provided by the integration  of operator-valued functions over a measure space.} 
In  Section  \ref{weyl} we derive the non-trivial unitary irreducible representation U of $\mathrm{H}^{\mathrm{dc}}_1$ {acting on the Hilbert space  $L^2(\mathbb{S}^1,\ud\gamma)$ of functions that are square  integrable on the circle.} This representation is square integrable on the {discrete cylinder}. We give an overview of the properties of the  {unitary} Weyl operator acting on $L^2(\mathbb{S}^1,\ud\gamma)$. We define the {related Gabor transform} associated with a {resolution of the identity and the resulting  family of coherent states on the circle}.
In Section  \ref{quant} we first define our quantization tools, namely a weight function $\varpi$ defined on the {discrete cylinder} and the related integral operator $\sfMv $ acting on the representation space. We then define in Section \ref{genform} the quantization map which associates a function $f $ on the {discrete cylinder} with an operator $A^{\varpi}_f$ acting in $L^2(\mathbb{S}^1,\ud\gamma)$. {We compute $A^{\varpi}_f$ for separable functions} in position and momentum, in momentum only, and position only. Finally, we examine smooth de-quantizations under the form of the {so-called}  semi-classical portraits $\check{f}$ of $A^{\varpi}_f$. 
In Section  \ref{portrait} we give some concrete examples of quantum operators built through coherent states for $\mathrm{H}^{\mathrm{dc}}_1$, and those yielded by the  weight $\varpi= 1$ corresponding to the parity operator.  
In Section \ref{weight}  we examine quantizations resulting from various choices of weight function $\varpi$. We conclude in Section \ref{conclu} 
with a few comments on the content of our work and on new directions to be explored. 
In Appendix \ref{table} {some  fiducial vectors and their corresponding  reproducing kernels are presented in a table}. 

\section{(Covariant) integral quantization: a survey}
\label{survey}

 Let $(X,\mu)$ be a measure space and $\mathcal{H}$ be a (separable) Hilbert space.   An operator-valued function
\begin{equation}
\label{}
X\ni x \mapsto \mathsf{M}(x)\ \mbox{acting in} \ \mathcal{H}\, , 
\end{equation}
resolves the identity operator $\mathbbm{1}$ in  $\mathcal{H}$ with respect to the measure $\mu$ if 
\begin{equation}
\label{resUn}
\int_X \mathsf{M}(x)\, \mathrm{d} \mu(x)= \mathbbm{1} 
\end{equation}
holds in a weak sense.

Integral quantization based on Eq.\;\eqref{resUn} is   the linear map of a function on $X$ to an operator in $\mathcal{H}$ which is defined by
\begin{equation}
\label{intqgen}
f(x) \mapsto \int_X f(x) \mathsf{M}(x)\, \mathrm{d} \mu(x):= A_f\, , \quad 1 \mapsto \mathbbm{1}\,.
\end{equation}
If the operators $\mathsf{M}(x)$ in Eq.\;\eqref{resUn}
are nonnegative and bounded,  one says that  they form a (normalised) positive operator-valued measure (POVM) on $X$.  If they are further
unit trace-class  for all $x\in X$, i.e., if the $\mathsf{M}(x)$'s are density operators, then the map 
\begin{equation}
\label{sempor}
f(x) \mapsto  \check{f}(x):= \mathrm{tr}(\mathsf{M}(x)A_f) = \int_X f(x^{\prime})\,\mathrm{tr}(\mathsf{M}(x)\mathsf{M}(x^{\prime}))\, \mathrm{d} \mu(x^{\prime})
\end{equation}
is a local averaging of the original $f(x)$ (which can very singular, like a Dirac) with respect to the probability distribution on $X$,
\begin{equation}
x^{\prime} \mapsto \mathrm{tr}(\mathsf{M}(x)\mathsf{M}(x^{\prime}))\,. 
\end{equation}
This averaging, or semi-classical portrait of the operator $A_f$,  is in general a regularisation, depending of course on the topological nature of the measure space $(X,\mu)$ and the functional properties of the $\mathsf{M}(x)$'s.

Now,  consider a set of parameters $\boldsymbol{\kappa}$   and corresponding families of  POVM  $ \mathsf{M}_{\boldsymbol{\kappa}}(x)$ solving the identity
\begin{equation}
\label{resunH}
\int_X \mathsf{M}_{\boldsymbol{\kappa}}(x)\, \mathrm{d} \mu(x)= \mathbbm{1}\, ,  
\end{equation}
 One says that the \textit{classical limit} $f(x)$ holds at $\boldsymbol{\kappa}_0$ if 
\begin{equation}
\check{f}_{{\boldsymbol{\kappa}}}(x):= \int_X f(x^{\prime})\,\mathrm{tr}(\mathsf{M}_{\boldsymbol{\kappa}}(x)\mathsf{M}_{\boldsymbol{\kappa}}(x^{\prime}))\, \mathrm{d} \mu(x^{\prime}) \to f(x) \quad \mbox{as} \quad \boldsymbol{\kappa} \to \boldsymbol{\kappa}_0\, , 
\end{equation}
where the convergence $\check{f}\to f$ is defined in the sense of a certain topology. 

 Otherwise said, $\mathrm{tr}(\mathsf{M}_{\boldsymbol{\kappa}}(x)\mathsf{M}_{\boldsymbol{\kappa}}(x^{\prime}))$ tends to  
\begin{equation}
\mathrm{tr}(\mathsf{M}_{\boldsymbol{\kappa}}(x)\mathsf{M}_{\boldsymbol{\kappa}}(x^{\prime})) \to \delta_x(x^{\prime})
\end{equation}
where $\delta_x$ is a Dirac measure with respect to $\mu$,
\begin{equation}
\int_X f(x^{\prime}) \, \delta_x(x^{\prime})\, \mathrm{d} \mu(x^{\prime}) = f(x)\,.
\end{equation}
Actually,  nothing guarantees the existence of such  a limit on a general level. Nevertheless if the semi-classical $\check{f}_{{\boldsymbol{\kappa}}}$ might appear as  more realistic and more easily manageable than the original $f$, the question is to evaluate  the range of acceptability of the parameters $\boldsymbol{\kappa}$.

Let us now assume  that $X=G$ is a Lie group with left Haar measure ${\rm d}\mu(g)$, and let $g \mapsto U(g)$ be a unitary irreducible representation (UIR) of $G$ in a Hilbert space $\mathcal{H}$. Let $\mathrm{M}$ be a bounded self-adjoint operator on $\mathcal{H}$ and let us define $g$-translations of $M$ as
\begin{equation}
\label{eqMg}
\mathrm{M}(g)= U(g) \mathrm{M} U(g)^\dagger\,.
\end{equation}
Suppose that the  operator
\begin{equation}
\label{intgrR}
R:= \int_G  \, \mathrm{M}(g)\,\ud\mu(g) \, ,
\end{equation}
is defined in a weak sense. From the left invariance of $\ud\mu(g)$  the operator $R$ commutes with all operators $U(g)$, $g\in G$, and so, from Schur's Lemma, we have the resolution of the unity up to a constant,
\begin{equation}
\label{resunitG}
R= c_{ \mathrm{M}}\mathbbm{1} \,.
\end{equation}
The constant $c_{ \mathrm{M}}$ can be found from the formula
\begin{equation}
\label{calcrho}
c_{ \mathrm{M}} = \int_G  \, \mathrm{tr}\left(\rho_0\, \mathrm{M}(g)\right)\, \ud\mu(g)\, ,
\end{equation}
where $\rho_0$ is a given unit trace positive operator. $\rho_0$ is  chosen, if manageable,  in order to make the integral convergent.  Of course, it is possible that no such finite constant exists for a given $\mathrm{M}$, or worse, it can not exist for any $\mathrm{M}$ (which is not the case for square integrable representations).
Now, if $c_{ \mathrm{M}}$ is finite and positive, the true resolution of the identity follows:
\begin{equation}
\label{Resunityrho}
\int_G \, \mathrm{M}(g) \,\ud \nu(g) = \mathbbm{1}\,, \quad \ud \nu(g):= \ud\mu(g)/c_{ \mathrm{M}}\, .
\end{equation}
For instance, in the case of a square-integrable unitary irreducible representation $U: g \mapsto U(g)$, let us pick a unit vector $| \psi \rangle$ for which $c_{\mathrm{M}} = \int_G {\rm d}\mu(g) |\langle \psi | U(g) \psi \rangle |^2 < \infty$, i.e $| \psi \rangle$ is an admissible unit vector for $U$. With $\mathrm{M} = |\psi \rangle \langle \psi |$ the resolution of the identity (\ref{Resunityrho}) provided by the family of states 
$| \psi_g \rangle = U(g) | \psi \rangle$ reads
\begin{equation}
\label{resunpsipsi}
 \int_G |\psi_g \rangle \langle \psi_g | \frac{{\rm d} \mu(g)}{c_{\mathrm{M}}} =\mathbbm{1}   \,.
 \end{equation}
Vectors $| \psi_g \rangle$ are named (generalized) coherent states for the group $G$. \\
 The equation (\ref{Resunityrho}) provides an integral quantization of complex-valued functions on the group $G$ as follows
 \begin{equation}
 \label{quantiz}
 f \mapsto A_f = \int_G \mathrm{M}(g) f(g) \frac{{\rm d} \mu(g)}{c_{\mathrm{M}}} \,.
 \end{equation}
 Furthermore, this quantization is \emph{covariant} in the sense that:
 \begin{equation}
 \label{covar} 
 U(g) A_f U(g)^\dagger = A_F \quad \text{where} \quad F(g') = (\mathfrak{U}(g) f)(g') = f(g^{-1} g')\,,
 \end{equation}
  i.e., $\mathfrak{U}(g): f \mapsto F$ is the left regular representation if $f \in L^2(G, {\rm d}\mu(g))$.

For the group $\mathrm{H}^{\mathrm{dc}}_1$ considered in this paper, an adaption of this  material   is necessary  in the sense that we have to replace {the group $\mathrm{H}^{\mathrm{dc}}_1$ with its right coset $\Gamma= \R\backslash\mathrm{H}^{\mathrm{dc}}_1\sim\Z\times \mathrm{SO}(2)$}, which amounts to  replace its UIR with its projective version. 

\section{Overview:  Weyl operator acting on $L^2(\mathbb{S}^1,\ud\gamma)$}
\label{weyl}
The group  $\mathrm{H}^{\mathrm{dc}}_1$ is the set of triplets $(s, m,\theta)\in \R\times\Z\times \mathbb{S}^1$ equipped with the following Weyl-Heisenberg-like multiplication:
\begin{equation}
(s,m,\theta)\,(s^{\prime}, m^{\prime},\theta^{\prime})= \left(s+s^{\prime} +\frac{m\theta^{\prime}-m^{\prime}\theta}{2},m+m^{\prime},\theta+\theta^{\prime} \, \mathrm{mod}\,2\pi\right)\,. 
\end{equation}
The neutral element is $e=(0,0,0)$ and the inverse is given by $$(s,m,\theta)^{-1}= (-s,-m,-\theta\,  \mathrm{mod}\,2\pi)\,. $$
{Let us define the  representation of $\mathrm{H}^{\mathrm{dc}}_1$ acting on $L^2(\mathbb{S}^1,\ud\gamma)$ as:
\begin{equation}
\label{repV}
({V}(s,m,\theta)\psi)(\gamma)=e^{\ii\,s}e^{-\ii\,\frac{m\theta}{2}}e^{\ii\,m\gamma}\psi(\gamma-\theta)\,.
\end{equation}
This representation is unitary and irreducible. 
Due to the central parameter $s\in \R$,  it is not square integrable on the group equipped with its semi-discrete Haar measure $\ud(s,m,g)$, i.e., there is no $\psi\in L^2(\mathbb{S}^1,\ud\gamma)$ such that $\sum_{m\in \Z}\int_{\R\times \mathbb{S}^1}\ud s\,\ud \theta\,\vert(\psi,({V}(s,m,\theta)\psi)\vert^2 < \infty$. On the other hand  it is square integrable on the right coset $\R\backslash\mathrm{H}^{\mathrm{dc}}_1=$ of $\mathrm{H}^{\mathrm{dc}}_1$ with its center. This right coset is identified with the phase space or discrete cylinder  $\Gamma=\Z\times\,\mathbb{S}^1$}. 
The representation \eqref{repV} induces the  unitary ($\sim$ Weyl) operator $U$ defined on  $\Gamma$ by the section $s=0$ on the group, i.e., ${U}(m,\theta):= {V}(0,m,\theta)$:
\begin{equation}
\label{UIR}
({U}(m,\theta)\psi)(\gamma)=e^{-\ii\,\frac{m\theta}{2}}\,e^{\ii\,m\gamma}\psi(\gamma-\theta)\,.
\end{equation}
One can show, through the use of the Schur's Lemma,  that for any $\phi$ in $L^2(\mathbb{S}^1,\ud\gamma)$ the family $\{{U}(m,\theta)\phi\}_{(m,\theta)\in\Gamma}$ constitutes an overcomplete  family 
resolving the identity in the sense of Eqs.\;\eqref{resUn} or \eqref{Resunityrho} or \eqref{resunpsipsi}:
\begin{equation}
\label{resunZS}
 \sum_{m\in\Z}\int_{\mathbb{S}^1}\ud\theta\, |{U}(m,\theta)\phi\rangle\langle{U}(m,\theta)\phi| = \mathbbm{1}\, .
\end{equation}
{ This implies that for any $\phi$, $\psi$ in $L^2(\mathbb{S}^1,\ud\gamma)$ we have:
\begin{equation}
\label{resunZSpp}
\sum_{m\in\Z}\int_{\mathbb{S}^1}\ud\theta\,\vert{\langle \psi|{U}(m,\theta)\phi\rangle\vert}^2= {\vert\vert\psi\vert\vert}^2 {\vert\vert\phi\vert\vert}^2 \,. 
\end{equation}}
{The function  $\Gamma \ni (m,\theta)\mapsto \phi_{(m,\theta)}:={U}(m,\theta)\phi$} is a coherent state (CS) for the group $\mathrm{H}^{\mathrm{dc}}_1$ in the sense given by  Perelomov \cite{perel86}, and the function $\phi$ is then called a fiducial vector.
The interpretation of the projection  of $\psi$  on $\phi_{(m,\theta)}$, namely
\begin{equation}
\langle\,\phi_{(m,\theta)}\vert\psi\rangle
=\int_{\mathbb{S}^1}\ud\gamma\, \overline{(U(m,\theta)\phi)(\gamma)}\psi(\gamma)
= \int_{\mathbb{S}^1}\ud\gamma\, e^{\ii\frac{m\theta}{2}}e^{-\ii\,m\gamma}\overline{\phi(\gamma-\theta)}\psi(\gamma)\, , 
\end{equation}
is the  phase space or momentum-angular position representation of $\psi$ with respect to the family of  coherent states $\phi_{(m,\theta)}$.

Let us now summarize the most important features of the above CS analysis of the elements of $L^2(\mathbb{S}^1,\ud\gamma)$ resulting from the resolution of the identity \eqref{resunZS}.
\beprop
\label{main_properties} {Let $L^2(\Gamma, \ud(m,\theta))$ be  the Hilbert space of square integrable functions on $\Gamma$ equipped with its semi-discrete  measure.} 
For any function $\phi, \psi\in\,L^2(\mathbb{S}^1,\ud\gamma)$, such that $\vert\vert\phi\vert\vert=1$, the map 
{\begin{equation}
L^2(\Gamma,\ud \gamma)\ni\psi\rightarrow\langle\phi_{(m,\theta)}\vert\psi\rangle \in L^2(\Gamma, \ud(m,\theta))
\end{equation} } satisfies the following properties:
\begin{itemize}
\item it is an isometry:
\begin{align}
\Vert\psi\Vert^2=\int_{\mathbb{S}^1}\ud\gamma\,\vert\psi(\gamma)\vert^2= \sum_{m\in\Z}\int_{\mathbb{S}^1}\ud\theta\,\vert{\langle\phi_{(m,\theta)}\vert\psi\rangle\vert}^2\,, 
\end{align}
\item it can be inverted on its range:
\begin{equation}
\psi(\gamma)= \sum_{m\in\Z}\int_{\mathbb{S}^1}\ud\theta\,\langle\phi_{(m,\theta)}\vert\psi\rangle\phi_{(m,\theta)}(\gamma)\, , 
\end{equation}
\item its range is a reproducing kernel space:
\begin{equation}
\Psi(m,\theta):=\langle\phi_{(m,\theta)}\vert\psi\rangle= \sum_{m^{\prime}\in\Z}\int_{\mathbb{S}^1}\ud\theta^{\prime}\, 
\langle\phi_{(m,\theta)}\vert\phi_{(m^{\prime},\theta^{\prime})}\rangle\Psi(m^{\prime},\theta^{\prime})\,,
\end{equation}
\end{itemize}
where the reproducing kernel $K_{\phi}$ is the two-point function on the phase space:
\begin{equation}
K_{\phi}((m,\theta), (m^{\prime},\theta^{\prime}))=\langle\phi_{(m,\theta)}\vert\phi_{(m^{\prime},\theta^{\prime})}\rangle\,.
\end{equation}
\enprop
On can show that:
\begin{equation}
K_{\phi}((m,\theta), (m^{\prime},\theta^{\prime}))=A(m,m^{\prime},\theta,\theta^{\prime})\Phi(m-m^{\prime},\theta-\theta^{\prime})
\end{equation}
\begin{equation}
A(m,m^{\prime},\theta,\theta^{\prime})= e^{\ii\,\frac{m(\theta-\theta^{\prime})-(m-m^{\prime})\theta^{\prime}}{2}}
\end{equation}
\begin{equation}
\Phi(m-m^{\prime},\theta-\theta^{\prime})= \int_{\mathbb{S}^1}\ud\gamma\,e^{-\ii(m-m^{\prime})}\overline{\phi(\gamma-(\theta-\theta^{\prime}))}\phi(\gamma)
\end{equation}

Let us now give a list of useful formulae for the sequel. 
\begin{equation}
\label{invU}
\begin{split}
U^{\dag}(m,\theta)\psi(\gamma)& = e^{\ii\frac{m\theta}{2}}e^{-\ii\,m(\gamma+\theta)}\psi(\gamma+\theta)
=e^{-\ii\frac{m\theta}{2}}e^{-\ii\,m\gamma}\psi(\gamma+\theta)\\ &=U(-m,-\theta)\psi(\gamma)\,.
\end{split}
\end{equation}
\begin{equation}
\label{UUp}
{U}(m,\theta)\,{U}(m^{\prime},\theta^{\prime})= e^{-\ii\frac{m\theta^{\prime}-m^{\prime}\theta} {2}}\,{U}(m+m^{\prime},\theta+\theta^{\prime})\,. 
\end{equation}
\begin{equation}
\label{UUUp}
{U}(m^{\prime},\theta^{\prime}){U}(m,\theta)\,{U}^{\dag}(m^{\prime},\theta^{\prime})= e^{\ii(m\theta^{\prime}-m^{\prime}\theta)}U(m,\theta)\,. 
\end{equation}
\begin{equation}
\label{UdUdp}
{U}^{\dag}(m,\theta)\,{U}^{\dag}(m^{\prime},\theta^{\prime})= e^{-\ii\frac{m\theta^{\prime}-m^{\prime}\theta} {2}}\,{U}^{\dag}(m+m^{\prime},\theta+\theta^{\prime})\,. 
\end{equation}
\begin{equation}
\label{TrU}
\mathrm{Tr}[U(m,\theta)]=\delta_{m\,0}\delta(\theta)\,. 
\end{equation}

As was stated,  any square integrable function on the circle can be used as a starting function, i.e., a fiducial vector to build  a coherent state. However, one is mostly interested in considering fiducial vectors that have some ``good'' localization both in angular position and momentum. One way of doing this is to take the periodized of a square integrable function that has good localization properties on the real line (see \cite{Richard_AU}, page 146, for the equivalent of the uncertainty principle on the circle). For this purpose, we use  periodization combined with the Poisson summation formula (see \cite{Richard_AU}, pages 137-138).
So, let us pick a function $\phi$ that is {integrable} on the real line. The periodization operator Per is defined by
\begin{equation}
(\mathrm{Per}\,\phi)(x)=2\pi\sum_{n\in\Z}\,\phi(x+2\pi\,n)\, , 
\end{equation}
and the application of the Poisson summation formula  reads:
\begin{equation}
2\pi\sum_{n\in\Z}\,\phi(x+2\pi\,n)= \sum_{n\in\Z}\,\mathcal{F}[\phi](n)e^{\ii\,n\,x}
\end{equation}
where $\mathcal{F}[\phi]$ is the Fourier transform of $\phi$, which is defined with its inverse  by
\begin{equation}
\mathcal{F}[\phi](\omega)=  \frac{1}{2\pi}\int_\R \ud\,x\,e^{-\ii\,\omega\,x}\phi(x)\,, \quad \phi(x)=\int_\R \ud\,x\,e^{\ii\,x\,\omega}\mathcal{F}[\phi](\omega)\,.  
\end{equation}
Let us present some examples  of such periodizations \cite{Richard_AU} and other examples of fiducial vectors.
\begin{itemize}
\item\,Periodized  ($L^2$ normalized ) gaussian kernel and theta function.
\begin{equation}
\frac{2\pi}{{(2\pi)}^{\frac{1}{4}}}\frac{1}{\sqrt{\sigma}}\sum_{n\in\Z}\,e^{-\frac{(x+2\pi\,n)^2}{2\sigma^2}}
= {2\pi(2\pi)}^{-\frac{3}{2}}\sum_{n\in\Z}\,e^{\ii\,n\,x}e^{-\frac{\sigma^2}{2}n^2}
= {(2\pi)}^{-\frac{1}{2}}\theta\left(\frac{x}{2\pi},\frac{\sigma}{\sqrt{2\pi}}\right)\,,
\end{equation}
where $\vartheta_{3}$ is the  third Jacobi theta function:
\begin{equation}
\vartheta_{3}(x,s)= \sum_{n\in\Z}\,e^{\ii\,2\pi\,n\,x}\,e^{\ii\pi\,s\,n^2}\,. 
\end{equation}
A number of authors (\cite{Rigasa,Kowalski1}) have used variants of this function as a fiducial vector for coherent states on the circle.

The corresponding reproducing kernel for $\sigma=1$ is:
\begin{equation}
\begin{split}
&\langle\phi_{(m,\theta)}\vert\phi_{(m^{\prime},\theta^{\prime})}\rangle=\\
&=(2\pi)^2\,e^{\ii\,\frac{m(\theta-\theta^{\prime})-(m-m^{\prime})\theta^{\prime}}{2}}e^{-\frac{(m-m^{\prime})^2}{2}}\sum_{n\in\Z}e^{-n^2}e^{\ii\,n\,(\ii(m-m^{\prime})+(\theta-\theta^{\prime}))}\\
&=(2\pi)^2\,e^{\ii\,\frac{m(\theta-\theta^{\prime})-(m-m^{\prime})\theta^{\prime}}{2}}e^{-\frac{(m-m^{\prime})^2}{2}}
\vartheta\left(\frac{\ii(m-m^{\prime})+(\theta-\theta^{\prime})}{2\pi},\frac{\ii}{\sqrt{\pi}}\right)\,. 
\end{split}
\end{equation}

\item\,Periodized  ($L^2$ normalized ) Poisson kernel.
\begin{equation}
P_{\sigma}(x)= 2\sum_{n\in\Z}\frac{\sigma^{-1}}{(x+2\pi\,n)^2+\sigma^{-2}}=\frac{1-r^2}{1-2\,r\,\cos(x)+r^2}\,. 
\end{equation}
\item Dirichlet fiducial vector ($L^2$ normalized ).
\begin{equation}
D_{n}(\gamma)= \frac{1}{\sqrt{2\pi(n+1)}}\sum_{m=-n}^{m=+n}\,e^{\ii\,n\,\varphi}=\frac{1}{\sqrt{2\pi(n+1)}}
\times\frac{\sin(n+\frac{1}{2})(\gamma)}{\sin{\frac{\gamma}{2}}}\,. 
\end{equation}
The corresponding reproducing kernel is:
\begin{equation}
\langle\phi_{(m,\theta)}\vert\phi_{(m^{\prime},\theta^{\prime})}\rangle\,
=\frac{1}{2\pi(n+1)}\,e^{\ii\,\frac{m(\theta-\theta^{\prime})-(m-m^{\prime})\theta^{\prime}}{2}}\frac{\sin((n+\frac{1}{2})(\theta-\theta^{\prime}))}{\sin(\frac{\theta-\theta^{\prime}}{2})}\,. 
\end{equation}
\item Fej\'er fiducial vector.
\begin{align}
&F_{n}(\varphi) =\frac{D_{0} (\varphi)+D_{1}(\varphi)+...+D_{n}(\varphi)}{n+1}\\\nonumber
&=\frac{1}{n+1} \left[{\frac{\sin((\frac{n+1}{2})\varphi)}{\sin{\frac{1}{2}\varphi}}}\right]^2\,
=\sum_{k=-n}^{k=n}(1-\frac{\vert\,k\vert}{n+1})e^{\ii\,k\varphi}. \\\nonumber
\end{align}
\item Von  Mises fiducial vector  ($L^2$ normalized)
\begin{equation}
\phi(\gamma)=\frac{e^{\lambda\cos(\gamma)}}{\sqrt{2\pi\,I_{0}(2\lambda)}}\, , 
\end{equation}
where $I_0$ is the modified Bessel of the first kind of zero order. The corresponding reproducing kernel is:
\begin{equation}
\langle\phi_{(m,\theta)}\vert\phi_{(m^{\prime},\theta^{\prime})}\rangle\,=\frac{e^{\ii\frac{(m-m^{\prime})\theta-m\,(\theta-\theta^{\prime})}{2}}}{2\pi\,I_{0}(2\lambda)}\times\,I_{m-m^{\prime}}(2\lambda)\,\cos\left(\frac{\theta-\theta^{\prime}}{2}\right)\,. 
\end{equation}
\end{itemize}
This fiducial vector is related to the Von Mises kernel which is widely used in the area of circular data analysis. In the context of quantum mechanics on the circle, it was derived by Carruthers in 1968 \cite{Carruthers}, and several other authors, including Kowalski \cite{Kowalski2} and collaborators, have used it.


\section{Quantization operators and the quantization map}
\label{quant}
Following previous works \cite{gazmur16,gazkoimur20,gazkoimur22}, we pick a function $\varpi$, called weight (but not necessarily positive),  on the phase space $\Gamma$. We then define the  operator $\sfMv $ {acting in $L^2(\mathbb{S}^1,\ud\gamma)$} by
\begin{equation}
\label{affbop}
\sfMv =\frac{1}{2\pi}\sum_{m\in\Z}\int_{\mathbb{S}^1}\ud\theta\,\varpi(m,\theta)\,U(m,\theta)\, .
\end{equation}

\beprop 
\begin{enumerate}
With the assumption  that the weight $\varpi$ has been chosen such that the operator $\sfMv $ is bounded: 
  \item[(i)] The  operator $\sfMv $ is the integral operator:
\begin{equation}
(\sfMv \psi)(\gamma)=\int_{\mathbb{S}^1}\ud \gamma\,\mathcal{M}^{\varpi}(\gamma,\gamma^{\prime})\psi(\gamma^{\prime})\, , 
\end{equation}
where the kernel ${\mathcal M}^{\varpi}(\gamma,\gamma^{\prime})$ is given by:
\begin{equation}
{\mathcal M}^{\varpi}(\gamma, \gamma^{\prime})= \frac{1}{2\pi}\sum_{m\in\Z}\,\varpi(m,\gamma-\gamma^{\prime})\,e^{\ii\frac{m(\gamma+\gamma^{\prime})}{2}}
\equiv\frac{1}{2\pi}\widetilde{\varpi}_{p_{1}}\left(\frac{\gamma+\gamma^{\prime}}{2},\gamma-\gamma^{\prime}\right)\, . 
\end{equation}
Here, $\widetilde{\varpi}_{p_{1}}$ is the  inverse discrete Fourier transform of $\varpi$ with respect to the first variable.
 \item[(ii)]   The operator $\sfMv $ is symmetric if and only the weight satisfies:
\begin{align}
\sfMv ={\sfMv }^{\dag} \Leftrightarrow \overline{\varpi(-m,\gamma)}=\varpi(m,-\gamma)\,.
\end{align}
 \item[(iii)] {The operator $\sfMv $ is traceclass and its trace  is given by}
\begin{equation}
\mathrm{Tr}(\sfMv )=\varpi(0,0)\,.
\end{equation}
\end{enumerate}
\enprop

\bprf
\begin{enumerate}
  \item[(i)] 
The action of $\sfMv $ on $\psi$ is given by:
\begin{align*}
(\sfMv \psi)(\gamma) &= \int_{\mathbb{S}^1}\ud\gamma^{\prime}\left(\frac{1}{2\pi}\sum_{m\in\Z}\,\varpi(m,\gamma-\gamma^{\prime})e^{\ii\frac{m(\gamma+\gamma^{\prime})}{2}}\right)\psi(\gamma^{\prime})\\
&=\int_{\mathbb{S}^1}\ud\gamma^{\prime}{\mathcal M}^{\varpi}(\gamma,\gamma^{\prime})\psi(\gamma^{\prime})\,.
\end{align*}

 \item[(ii)]  The condition that $\sfMv $ be symmetric implies the following condition on the kernel:
\begin{equation*}
\overline{{\mathcal M}^{\varpi}(\gamma, \gamma^{\prime})}= {\mathcal M}^{\varpi}(\gamma^{\prime},\gamma)
\end{equation*}
which gives:
\begin{equation*}
\overline{\widetilde{\varpi}_{p_{1}}\left(\frac{\gamma+\gamma^{\prime}}{2},\gamma-\gamma^{\prime}\right)}=\widetilde{\varpi}_{p_{1}}\left(\frac{\gamma+\gamma^{\prime}}{2},-(\gamma-\gamma^{\prime}))\right).
\end{equation*}
 \item[(iii)] Therefore the trace of $\sfMv $ corresponds the integral of the kernel over its diagonal, that is:
\begin{equation*}
\int_{\mathbb{S}^1}\ud\gamma\,{\mathcal M}^{\varpi}(\gamma,\gamma)= \frac{1}{2\pi}\int_{\mathbb{S}^1}\ud\gamma\sum_{m\in\Z}\,\varpi(m,0)e^{\ii\,m\gamma}
=\frac{1}{2\pi}\sum_{m\in\Z}\,\varpi(m,0)2\pi\delta_{m\,0}=\varpi(0,0).
\end{equation*}
\end{enumerate}
\eprf
In turn, one retrieves the weight $\varpi$ from the quantization operator $\sfMv$ through a tracing operation.
\beprop
\label{tretrw}
The  trace of the operator ${U}^{\dag}(m,\theta)\sfMv $ is given by:
\begin{equation}
\label{UMvptr}
\mathrm{Tr}[U^{\dag}(m,\theta)\sfMv ]=\varpi(m,\theta)
\end{equation}
\enprop
\bprf
To compute this trace one uses the expansion  of the kernel  in terms of  the orthonormal {Fourier} basis $\left\{e_n(\gamma)= \frac{1}{\sqrt{2\pi}}e^{\ii\,n\gamma}\right\}$ as follows
\begin{align*}
&\mathrm{Tr}[U^{\dag}(m,\theta)\sfMv ] =\sum_{n\in\Z}\langle\,e_{n}\vert\,U^{\dag}(m,\theta)\sfMv e_{n}\rangle
=\sum_{n\in\Z}\langle\,U(m,\theta)\,e_{n}\vert\,\sfMv e_{n}\rangle\\
&=\sum_{n\in\Z}\int_{\mathbb{S}^1}d\gamma\frac{1}{\sqrt{2\pi}}\, e^{\ii\frac{m\theta}{2}}\,e^{-\ii\,m\gamma}\,e^{-\ii\,n\,(\gamma-\theta)}\frac{1}{2\pi}\sum_{m^{\prime}\in\Z}\int_{\mathbb{S}^1}d\theta^{\prime}\,\varpi(m^{\prime},\theta^{\prime})e^{-\ii\frac{m^{\prime}\theta^{\prime}}{2}}e^{\ii\,m^{\prime}\gamma}\frac{1}{\sqrt{2\pi}}\,e^{\ii\,n(\gamma-\theta^{\prime})}\\
&=\frac{1}{(2\pi)^2}\sum_{m^{\prime}\in\Z}\int_{\mathbb{S}^1}d\theta^{\prime}\,\varpi(m^{\prime},\theta^{\prime}) 2\pi\delta(\theta-\theta^{\prime})2\pi\delta_{m\,m^{\prime}}\\
&=\varpi(m,\theta).
\nonumber
\end{align*}
\eprf
\section{Covariant affine integral quantization from weight function}
\label{genform}
\subsection{General results}
\label{genqres}
We now  establish general formulae for the integral quantization issued from 
a weight function $\vap(m ,\theta )$ on $\Gamma=\Z\times \mathbb{S}^1$ yielding the bounded self-adjoint operator $\sfMv$ defined in \eqref{affbop}. 
This allows us to build a family of operators obtained from the Weyl operator transport of $\sfMv$:
\begin{equation}
\label{WeylMom}
\sfMv(m,\theta )=U( m , \theta )\sfMv U^{\dag}( m , \theta )\,. 
\end{equation}
Then, the corresponding integral quantization is given by the linear map:
\begin{equation}
\label{genqvap}
f\mapsto A^{\vap}_f= \frac{1}{2\pi}\sum_{\Z}\int_{\mathbb{S}^1}\ud\gamma\, f(m,\theta )\, \sfMv(m,\theta)\,.
\end{equation}
We  have the following result.
\beprop
 $A^{\varpi}_f$ is the integral operator on $L^2(\mathbb{S}^1,\ud\gamma)$
\begin{equation}
(A^{\varpi}_f\psi)(\gamma)=\int_{\mathbb{S}^1} \ud\gamma^{\prime} {\mathcal A}^{\varpi}(\gamma, \gamma^{\prime}) \psi(\gamma^{\prime})\,, 
\end{equation}
and its  kernel is given by
\begin{align}
\label{kernel_A_f}
\mathcal{A}_{f}^{\varpi}(\gamma, \gamma^{\prime}) = \frac{1}{2\pi} \sum_{n\in\Z} F(\gamma-\gamma^{\prime},n)\varpi(n,\gamma-\gamma^{\prime})e^{\ii\,n\frac{(\gamma+\gamma^{\prime})}{2}}\,,
\end{align}
where $F(\varphi,k)$ is the  Fourier transform of $f$ with respect to the angle variable and the inverse (discrete) Fourier transform in the momentum variable,
\begin{align}
\label{fourier_two_var}
F(\varphi,n)= \frac{1}{2\pi} \sum_{k\in\Z}\int_{\mathbb{S}^1}\ud\theta\, f(k,\theta )e^{\ii (k\varphi-n\theta)}
\end{align}
\enprop
\bprf
The calculation of the kernel of the integral operator $A^{\vap}_f$ goes through the following steps.
\begin{align*}
&(A^{\vap}_f)(\psi)(\gamma)
= \frac{1}{2\pi}\sum_{m\in\Z}\int_{\mathbb{S}^1}\ud\theta\, f(m,\theta )\, \left(\sfMv(m,\theta) \psi\right)(\gamma)\\
&=\frac{1}{(2\pi)^2} \sum_{m\in\Z}\sum_{n\in\Z}\int_{\mathbb{S}^1}\ud\theta\int_{\mathbb{S}^1}\ud\alpha\, f(m,\theta )\varpi(n,\alpha)\, \left(U(m,\theta)U(n,\alpha)U^{\dag}(m,\theta) \psi\right)(\gamma)\\\nonumber
&=\frac{1}{(2\pi)^2} \sum_{m\in\Z}\sum_{n\in\Z}\int_{\mathbb{S}^1}\ud\theta\int_{\mathbb{S}^1}\ud\alpha\, f(m,\theta )\varpi(n,\alpha)e^{\ii(m\alpha-n\theta)}\left(U(n,\alpha) \psi\right)(\gamma)\\\nonumber
&=\frac{1}{(2\pi)^2}  \sum_{m\in\Z}\sum_{n\in\Z}\int_{\mathbb{S}^1}\ud\theta\int_{\mathbb{S}^1}\ud\alpha\, f(m,\theta )\varpi(n,\alpha)e^{\ii(m\alpha-n\theta)}e^{-\ii\frac{n\alpha}{2}}e^{\ii\,n\gamma}\psi(\gamma-\alpha)\\\nonumber
&=\frac{1}{(2\pi)^2}  \sum_{m\in\Z}\sum_{n\in\Z}\int_{\mathbb{S}^1}\ud\theta\int_{\mathbb{S}^1}\ud\gamma^{\prime}\, f(m,\theta )\varpi(n,\gamma-\gamma^{\prime})e^{\ii(m(\gamma-\gamma^{\prime})-n\theta)}e^{-\ii\frac{n(\gamma-\gamma^{\prime})}{2}}e^{\ii\,n\gamma}\psi(\gamma^{\prime})\\\nonumber
&\equiv\int_{\mathbb{S}^1}\ud\gamma^{\prime}\left\{\frac{1}{2\pi} \sum_{n\in\Z} F(\gamma-\gamma^{\prime},n)\varpi(n,\gamma-\gamma^{\prime})e^{\ii\,n\frac{(\gamma+\gamma^{\prime})}{2}}\right\}\psi(\gamma^{\prime})
\end{align*}
\eprf
One easily checks that $f=1$ gives $F(\varphi,k)=2\pi\delta(\varphi)\delta_{k\,0}$. Inserting this in $\eqref{kernel_A_f}$ gives:
\begin{equation}
\label{kernel_A_1}
\mathcal A_{1}^{\varpi}(\gamma, \gamma^{\prime})
= \frac{1}{2\pi}\sum_{n\in\Z} 2\pi\delta(\gamma-\gamma^{\prime})
\delta_{n\,0}\varpi(n,\gamma-\gamma^{\prime})\,e^{\ii\,n\frac{(\gamma+\gamma^{\prime})}{2}}
= \delta(\gamma-\gamma^{\prime})\varpi(0,0)
= \delta(\gamma-\gamma^{\prime})\\
\end{equation}
for $\varpi(0,0)=1$.

Covariance in the sense given by Eq. \eqref{covar}  is easily proven in the present case.
\beprop
The quantization map $f\mapsto A^{\vap}_f$ is covariant with respect to the action of the representation $V$, that is,
\begin{equation}
\label{genqvap}
V\,A^{\vap}_f V^{\dag}= \,A^{\vap}_{\mathcal{V}f}
\end{equation}
where $(\mathcal{V}(s,m,\theta)f)(n,\varphi)= f(n-m,\varphi-\theta)$ is the induced action  on the phase space.
\enprop

Let us summarize the above set of results. 
\beprop
\label{weightquant}
Let $\vap(m ,\theta )$ be a weight function  on $\Gamma=\Z\times \mathbb{S}^1$ obeying 
\begin{equation}
\label{weightcond}
\varpi(0,0)=1\ , \quad \overline{\varpi(-m,\gamma)}=\varpi(m,-\gamma)\, , 
\end{equation}
and such  that the operator $\sfMv$ defined in \eqref{affbop} is bounded. It results that the  operators $\sfMv(m,\theta )=U( m , \theta )\sfMv U^{\dag}( m , \theta )$ are bounded self-adjoint, resolve the identity
\begin{equation}
\label{residMotr}
\frac{1}{2\pi}\sum_{\Z}\int_{\mathbb{S}^1}\ud\theta\, \sfMv(m,\theta)=\mathbbm{1}
\end{equation}
and yield a covariant integral quantization of functions (or distributions) on the phase space $\Gamma$. 
\enprop

In the sequel we suppose that the choice of a weight function complies with the conditions of Proposition \ref{weightquant}.

\subsection{Quantization of separable functions}
As a first application, we consider the quantization of a separable function:
\begin{equation}
f(k,\theta)=g(k)\,h(\theta)\,.
\end{equation}
In this case the function $F$ in \eqref{kernel_A_f} and the integral kernel read as 
\begin{equation}
\label{Fgh}
F(\gamma-\gamma^{\prime},n)= \frac{1}{2\pi} \sum_{k\in\Z}\int_{\mathbb{S}^1}\ud\theta\, g(k)h(\theta)e^{\ii (k(\gamma-\gamma^{\prime})-n\theta)}
= \tilde{g}(\gamma-\gamma^{\prime})\hat{h}(n)\,,
\end{equation}
\begin{equation}
\label{A_f_separab}
\mathcal{A}_{gh}^{\varpi}(\gamma, \gamma^{\prime}) = \tilde{g}(\gamma-\gamma^{\prime}) \frac{1}{2\pi}  \sum_{n\in\Z} \hat{h}(n)\varpi(n,\gamma-\gamma^{\prime})e^{\ii\,n\frac{(\gamma+\gamma^{\prime})}{2}}\,.
\end{equation}

\subsection{Quantization of functions of momentum only}
The restriction  of \eqref{Fgh} and \eqref{A_f_separab} to a function of momentum only, $f(m,\theta)=g(m)$, yields for the function $F$ and the corresponding integral kernel:
\begin{equation}
F(\gamma-\gamma^{\prime},-n)=\tilde{g}(\gamma-\gamma^{\prime})\delta_{n\,0}\,, 
\end{equation}
\begin{equation}
\label{kernel_A_{g}}
\begin{split}
\mathcal{A}_{g}^{\varpi}(\gamma, \gamma^{\prime})& =  \frac{1}{2\pi}\sum_{n\in\Z} \tilde{g}(\gamma-\gamma^{\prime})\delta_{n\,0}\varpi(n,\gamma-\gamma^{\prime})e^{-\ii\,n\frac{(\gamma-\gamma^{\prime})}{2}}\\
&= \frac{1}{2\pi}\tilde{g}(\gamma-\gamma^{\prime})\varpi(0,\gamma-\gamma^{\prime})\,, 
\end{split}
\end{equation}
where $\tilde{g}(\gamma)=\sum_{m\in\Z}g(m)e^{\ii\,m\gamma}$.
It results the following action of the operator $A^{\varpi}_g$:
\begin{equation}
\label{Afh}
(A^{\varpi}_g\psi)(\gamma)= \frac{1}{2\pi}\int_{\mathbb{S}^1}\ud\gamma^{\prime} \tilde{g}(\gamma^{\prime}-\gamma)\varpi(0,\gamma-\gamma^{\prime}) \psi(\gamma^{\prime})\,. 
\end{equation}
Let us give  some elementary examples:
\begin{itemize}
\item Angular momentum $g(m)=m$, through integrating by parts {and appropriate derivability properties of the weight function}, 
\begin{align*}
(A^{\varpi}_m\psi)(\gamma)
&= \frac{1}{2\pi}\sum_{m\in\Z}\int_{\mathbb{S}^1}\ud\gamma^{\prime} \left(\frac{\partial}{-\ii\partial\gamma^{\prime}}{e^{\ii\,m(\gamma-\gamma^{\prime})}}\right)\varpi(0,\gamma-\gamma^{\prime}) \psi(\gamma^{\prime})\\\nonumber
&= \frac{1}{2\pi}\sum_{m\in\Z}\int_{\mathbb{S}^1}\ud\gamma^{\prime} e^{\ii\,m(\gamma-\gamma^{\prime})}
\left\{-\ii\frac{\partial}{\partial\gamma^{\prime}}-\frac{\partial}{\partial\gamma^{\prime}}[\ii\varpi(0,\gamma-\gamma^{\prime})]\right\} \psi(\gamma^{\prime})\\\nonumber
&=\left(-\varpi(0,0)\,\ii\frac{\partial}{\partial\gamma}-\frac{\partial}{\partial\gamma}[\ii\varpi(0,\gamma)]_{\gamma=0}\right) \psi(\gamma)= \left(-\ii\frac{\partial}{\partial\gamma}-\frac{\partial}{\partial\gamma}[\ii\varpi(0,\gamma)]_{\gamma=0}\right) \psi(\gamma)\, ,
\end{align*}
and so, with  the definition 
\begin{equation}
\label{Omdef}
\Omega(\gamma):=\varpi(0,\gamma)\, , \quad \Omega(0)=1\, ,
\end{equation}\
\begin{equation}
\label{A_{m}}
A^{\varpi}_m =-\ii\frac{\partial}{\partial\gamma}-\frac{\partial}{\partial\gamma}[\ii\varpi(0,\gamma)]_{\gamma=0}\equiv L-\Omega^{\prime}(0)\,. 
\end{equation}
Hence with a weight function obeying $\varpi(0,\gamma)]_{\gamma=0}= \Omega^{\prime}(0)=0$ we retrieve the usual  angular momentum operator $L=-\ii\frac{\partial}{\partial\gamma}$.
\item Square angular momentum $g(m)=m^2$. Along similar methods {and assumptions on the weight function}, we find
\begin{align*}
(A^{\varpi}_{m^2}\psi)(\gamma)&= \frac{1}{2\pi}\sum_{m\in\Z}\int_{\mathbb{S}^1}\ud\gamma^{\prime} \left(-{\frac{\partial^2}{\partial\gamma{^{\prime\,2}}}}{e^{\ii\,m(\gamma-\gamma^{\prime})}}\right)\varpi(0,\gamma-\gamma^{\prime}) \psi(\gamma^{\prime})\\
\nonumber
&= \frac{1}{2\pi}\sum_{m\in\Z}\int_{\mathbb{S}^1}d\gamma^{\prime} e^{\ii\,m((\gamma-\gamma^{\prime})}
\{-\frac{\partial^2}{\partial\gamma^{\prime\,2}}[\varpi(0,\gamma-\gamma^{\prime}) \psi(\gamma^{\prime})]\\\nonumber
&=\left(-\Omega(0)\frac{\partial^2}{\partial\gamma^{\,2}}+2\Omega^{\prime}(0)\frac{\partial}{\partial\,\gamma}-\Omega^{\prime\prime}(0)\right)\psi(\gamma)\,. 
\end{align*}
Finally, 
\begin{equation}
\label{A_{m^2}}
A^{\varpi}_{m^2}= L^2+2\ii\frac{\partial}{\partial\gamma}\Omega^{\prime}(0)\,L-\Omega^{\prime\prime}(0)\,.
\end{equation}
\end{itemize}
{\begin{remark}
At this point it is valuable to compare the above quantization of functions defined on the discrete cylinder $\Gamma = \Z\times \mathbb{S}^1$ with the quantization of the same functions defined on the cylinder $\mathcal{C}:=\R\times \mathbb{S}^1$, for instance that one using De Bi\`evre or  similar coherent states on the circle (see  \cite{Fresneda} and references therein). Both approaches yield  similar results (up to an additive constant) for the classical momentum, i.e., the angular momentum,  and its square, i.e., the kinetic energy, of a particle moving on the circle. Is it a kind of elementary illustration of Hamiltonian reduction in the sense of Marsden and Weinstein \cite{marsden74} ?
\end{remark}}

\subsection{Quantization of function of angular position only}
We now turn to the quantization of a function of the position only, $f(m,\theta)= h(\theta)$.
From
\begin{equation}
\tilde{f}(\gamma-\gamma^{\prime},n)= \delta(\gamma-\gamma^{\prime})\hat{h}(n)\, , 
\end{equation}
we obtain for the  integral kernel:
\begin{equation}
\label{kernel_A_h}
\mathcal{A}_{h}^{\varpi}(\gamma, \gamma^{\prime})
= 2\pi\delta(\gamma-\gamma^{\prime})\sum_{n\in\Z} \hat{h}(n)\, \varpi(n,\gamma-\gamma^{\prime})e^{\ii\,n\frac{(\gamma+\gamma^{\prime})}{2}}\,.
\end{equation}
This yields the multiplication operator 
{\begin{equation}
\label{A_f(theta)}
(A^{\varpi}_h\psi)(\gamma)= \left(\frac{1}{2\pi}\sum_{n\in\Z} \hat{h}(n)\, \varpi(n,0)e^{\ii\,n\gamma}\right)\psi(\gamma):= h^{\varpi}(\gamma)\psi(\gamma)\, . 
\end{equation}
  We observe that the coefficients of  Fourier series of the factor $h^{\varpi}(\gamma)$ are the those of the original $h$ multiplied by $\varpi(n,0)$.} 

\section{Semi-classical portraits}
\label{portrait}
Given a function $\varpi(m,\theta)$ on the phase space $\Gamma$, normalised at $\varpi(0,0)=1$, and yielding a positive unit trace operator, i.e. a density operator, $\sfMv $, the quantum phase space portrait of an operator $A$ on $L^2(\Gamma,\ud\gamma)$ reads as:
\begin{equation}
\label{semclA}
\check{A}(m,\theta):=  \mathrm{Tr}\left(A\,U(m,\theta)\,\sfMv U^{\dag}(m,\theta )\right) = \mathrm{Tr}\left(A\,\sfMv (m,\theta)\right)\, . 
\end{equation}
The most interesting aspect of this notion in terms of probabilistic interpretation holds when the operator $A$ is precisely the quantized version $A^\varpi_f$ of a classical $f(m,\theta)$ with the same function $\varpi$ (actually we could define the transform with two different ones, one for the ``analysis'' and the other for the ``reconstruction''). Then, with the  use of the composition rule \eqref{UUp}, we compute the transform:
\begin{align*}
f( m,\theta )&\mapsto \check f(m,\theta) \equiv \check{A}_f^{\vap}( m,\theta )=\mathrm{Tr}\left(A^{\varpi}_{f}\,\sfMv (m,\theta)\right)\\\nonumber
&=\frac{1}{2\pi}\sum_{m^{\prime}\in\Z}\int_{\mathbb{S}^1}\ud\theta^{\prime}\, f(m^{\prime},\theta^{\prime} )\mathrm{Tr}\left(U(m^{\prime},\theta^{\prime})\sfMv U^{\dag}(m^{\prime},\theta^{\prime})\,U(m,\theta)\sfMv U^{\dag}(m,\theta)\right)\\\nonumber
&=\frac{1}{2\pi}\sum_{m^{\prime}\in\Z}\int_{\mathbb{S}^1}\ud\theta^{\prime}\, f(m^{\prime},\theta^{\prime} )\mathrm{Tr}\left(\sfMv \,U(m-m^{\prime},\theta-\theta^{\prime})\sfMv U(-(m-m^{\prime}),-(\theta-\theta^{\prime}))\right)\\\nonumber
&=\frac{1}{2\pi}\sum_{m^{\prime}\in\Z}\int_{\mathbb{S}^1}\ud\theta^{\prime}\, f(m-m^{\prime},\theta-\theta^{\prime} )\mathrm{Tr}\left(\sfMv \,U(m^{\prime},\theta^{\prime})\sfMv U^{\dag}(m^{\prime},\theta^{\prime})\right)\\\nonumber
&=\frac{1}{2\pi}\sum_{m^{\prime}\in\Z}\int_{\mathbb{S}^1}\ud\theta^{\prime}\, f(m-m^{\prime},\theta-\theta^{\prime} )\mathrm{Tr}\left(\sfMv\,\sfMv (m^{\prime},\theta^{\prime})\right)\,. 
\end{align*}
Finally we get:
\beprop
The semi-classical portrait of the operator $A^{\varpi}_{f}$ with respect to the weight $\varpi$ is given by:
\begin{equation}
\label{lowsymbv}
 \check f(m,\theta)
=\frac{1}{2\pi}\sum_{m^{\prime}\in\Z}\int_{\mathbb{S}^1}\ud\theta^{\prime}\, f(m-m^{\prime},\theta-\theta^{\prime} )\mathrm{Tr}\left(\sfMv (m^{\prime},\theta^{\prime})\,\sfMv\right)\,. 
\end{equation}
\enprop
This expression, which can be viewed as a convolution on the phase space, has the meaning of an averaging of the classical $f$. The function 
\begin{align}
\label{distvap}
 (m,\theta) &\mapsto \mathrm{Tr}\left(\sfMv(m,\theta)\sfMv\right)\, ,
\end{align}
is a true probability distribution on $\Gamma$, i.e., is positive and with integral on $\Gamma$ equal to 1,    from the resolution of the identity and the fact  that $\sfMv$ is chosen as a density operator. Expression \eqref{distvap} is actually a kind of Husimi function.
The  expression of $\mathrm{Tr}\left(\sfMv(m,\theta )\sfMv\right)$  is easily derived and reads as
\begin{equation}
\label{trmommom}
\begin{split}
\mathrm{Tr}\left(\sfMv(m,\theta )\sfMv\right)
&=\frac{1}{2\pi}\sum_{n\in\Z}\langle\,e_{n}\vert\left(\,\sfMv (m,\theta)\sfMv \right)\,e_{n}\rangle\\
&=\frac{1}{2\pi}\sum_{m^\prime\in \Z}\int_{\mathbb{S}^1}d\theta^{\prime}\vert\varpi(m^{\prime},\theta^{\prime})\vert^{2}e^{\ii\,(m\theta^{\prime}-m^{\prime}\theta)}\,. 
\end{split}
\end{equation}
Integrating this expression on $\Gamma$, we get $1$, which means that $\check 1= 1$, as expected.

\section{Quantization with various weights}
\label{weight}
\subsection{Weight related to  coherent states for $\mathrm{H}^{\mathrm{dc}}_1$}

The weight $\varpi_\psi$ corresponding to the projection operator $\vert\psi\rangle\langle\psi\vert$ of a square integrable function $\psi$, with unit norm,  on $\mathbb{S}^1$ is  found through  the trace formula given by Proposition \ref{tretrw}:
\begin{equation}
\label{weight_cs}
\varpi_{\psi}(m,\theta)= \mathrm{Tr}[U^{\dag}(m,\theta)\vert\psi\rangle\langle\psi\vert]= \langle U(m,\theta)\psi\vert\psi\rangle
\end{equation}
In this case, from $\varpi_\psi=\langle U(m,\theta)\psi\vert\psi\rangle$, and from the Fourier expansion of  $\psi\in L^2(\mathbb{S}^1,\ud\gamma)$,  
\begin{equation}
\label{fourpsi}
\psi = \sum_{m\in\Z} \hat{\psi}(m) e_m\, , \quad \hat{\psi}(m)= \lg e_m|\psi\rg= \int_{\mathbb{S}^1} \ud \gamma\,\frac{ e^{-\ii m \gamma}}{\sqrt{2\pi}}\, \psi(\gamma)\, ,
\end{equation}
we have for $\Omega_{\psi}(\gamma)= \varpi_\psi(0,\gamma)$: 
\begin{align}
&\Omega_{\psi}(\gamma)= \langle U(0,\gamma)\psi\vert\psi\rangle=\sum_{m\in\Z}e^{\ii\,m\gamma}\vert\hat{\psi}(m)\vert^2=\left\lg e^{\ii\,m\gamma}\right\rg_{\vert\hat\psi\vert^2}\,,\\
&\Omega_{\psi}^{\prime}(0)=\ii\sum_{m\in\Z}\,m\,\vert\hat{\psi}(m)\vert^2=\ii \lg m\rg_{\vert\hat\psi\vert^2}\,, \\
&\Omega_{\psi}^{\prime\prime}(0))=-\sum_{m\in\Z}\,m^2\vert\hat{\psi}(m)\vert^2 = -\lg m^2\rg_{\vert\hat\psi\vert^2}\,,
\end{align}
where $\lg\cdot\rg_{\vert\psi\vert^2}$ means the average of the random variable ``$\cdot$'' with respect to the discrete probability distribution  $\Z\ni m\mapsto \vert\hat{\psi}(m)\vert^2$ ($\psi$ has unit norm). 
We then derive  the following quantizations of the  angular momentum, of its square, and a function of the angle only.
\begin{itemize}
\item $g(m)=m$, and from $\eqref{A_{m}}$, we get:
\begin{equation}
A^{\varpi_{\psi}}_{m}=-\ii\frac{\partial}{\partial\,\gamma}+  \lg m\rg_{\vert\hat\psi\vert^2}= L +  \lg m\rg_{\vert\hat\psi\vert^2}\, . 
\end{equation}
\item $g(m)=m^2$ and from $\eqref{A_{m^2}}$, we get:
\begin{equation}
A^{\varpi_{\psi}}_{m^2}=L^2 +2\lg m\rg_{\vert\hat \psi\vert^2}L+ \lg m^2\rg_{\vert\hat\psi\vert^2}\,. 
\end{equation}
\item $h(\theta)$ and from $\eqref{weight_cs}$ we get the multiplication operator:
\begin{equation}
\label{A_f(theta)}
\begin{split}
\left(A^{\varpi_{\psi}}_h\phi\right)(\gamma)&= \left[\sum_{m\in\Z} \hat{h}(m)\, \langle\,U(m,0)\psi\vert\psi\rangle\,e^{\ii\,m\gamma}\right]\phi(\gamma)\\
&= \left[ \sum_{m\in\Z} \hat{h}(m)\,\lg e^{-\ii m\gamma}\rg_{\vert\psi\vert^2}\,e^{\ii\,m\gamma}\right]\phi(\gamma)\,.
\end{split}
\end{equation}
\end{itemize}
Note that for  the CS weight \eqref{weight_cs}, the expression \eqref{distvap} $\mathrm{Tr}\left(\sfMv(m,\theta )\sfMv\right)$ is the Fourier transform of the Husimi function associated with $\psi$.

\subsection{Weight related to the {angular} parity operator}

The weight $\varpi_{\sfP}$ corresponding to the parity operator $\sfP\psi(\gamma)= \psi(2\pi-\gamma)$ on a square integrable function $\psi$ on $\mathbb{S}^1$ is simply:
\begin{equation}
\label{varpi_parity}
\varpi_{\sfP}(m,\theta)= 1\,.
\end{equation}
\bprf
\begin{align*}
&2\,\mathrm{Tr}[U^{\dag}(m,\theta)\sfP]
=2\sum_{n\in\Z}\langle\,e_n | U(m,\theta)^{\dag}\sfP\,e_n\rangle\\
&=2\frac{1}{2\pi}\sum_{n\in\Z}\int_{\mathbb{S}^1}d\gamma e^{-\ii\,n\gamma}e^{-\ii\frac{m\theta}{2}}e^{-\ii\,m\gamma} e^{-\ii\,n(\theta+\gamma)}=2\frac{1}{2\pi}\sum_{n\in\Z}\int_{\mathbb{S}^1}d\gamma e^{-\ii\,m(\gamma+\frac{\theta}{2})} e^{-\ii\,n(\theta+2\gamma)}\\\nonumber
&=2\frac{1}{2\pi} \int_{\mathbb{S}^1}d\gamma\,e^{-\ii\,m(\gamma+\frac{\theta}{2})}2\pi\delta(2\gamma+\theta)
=1\,.
\end{align*}
\eprf
Defining a Wigner-like  distribution of $\psi$ on the phase $\Z\times\,\mathbb{S}^1$ as
\begin{equation}
\label{wignerpsi}
W_\psi(m,\theta) =  \frac{1}{2\pi}\int_{S^{1}} \ud\gamma\,e^{-\ii\,m\,\gamma}\overline{\psi\left(\theta-\frac{\gamma}{2}\right)}\psi\left(\theta+\frac{\gamma}{2}\right)\,, 
\end{equation}
let us see how it is linked to the parity operator $\sfP$ and its corresponding uniform weight $\varpi_{\sfP}(m,\theta)= 1$. It was derived my Mukunda in 1979 \cite{mukunda1}. It is interesting to compare the following result derived from the $\mathrm{H}^{\mathrm{dc}}_1$ symmetry with the one issued from  the Weyl-Heisenberg $\mathrm{H}_1$ symmetry \cite{grossmann}. 
\beprop
The Wigner-like  distribution of $\psi$ on the phase $\Z\times\,\mathbb{S}^1$ is  the mean value $\times 1/2\pi$ of the Weyl-Heisenberg transport of the parity operator in the state $\psi$ :
\begin{equation}
\label{ }
W_\psi(m,\theta) = \frac{1}{2\pi} \langle\psi| \sfP(m,\theta)\psi\rangle\, , \quad \mbox{with}\quad  \sfP(m,\theta) = U(m,\theta)\sfP U(m,\theta)^{\dag}\,. 
\end{equation}
\enprop
\bprf
\begin{align*}
W_\psi(m,\theta) &= \frac{1}{2\pi} \langle\psi\vert\,\sfP (m,\theta)\psi\rangle= \frac{1}{2\pi} \langle\psi\vert\,{U}(m,\theta)\, \sfP \, {U}^{\dag}(m,\theta)\psi\rangle\\\nonumber
 &=  \frac{1}{2\pi}\langle\,{U}^{\dag}(m,\theta))\psi\vert\, \sfP \, {U}^{\dag}(m,\theta)\psi\rangle\\
 &= \frac{1}{(2\pi)^2} \int_{\mathbb{S}^{1}}\ud\gamma\,e^{\ii\frac{m\theta}{2}}e^{\ii\,m\gamma}\overline{\psi(\gamma+\theta)} \times\\
&\times\sum_{n\in\Z}\int_{\mathbb{S}^{1}} \ud\varphi\,   e^{\ii\,n\,\gamma} e^{-\ii\frac{n\varphi}{2}}e^{-\ii\frac{m\theta}{2}}\,e^{-\ii\,m\,(\gamma-\varphi)} \psi(\gamma-\varphi+\theta) \\
 &= \frac{1}{(2\pi)^2}\int_{\mathbb{S}^{1}}\ud\gamma\,\int_{\mathbb{S}^{1}} d\varphi\,e^{\ii\,m\,\varphi}\overline{\psi(\gamma+\theta)}
\psi(\gamma-\varphi+\theta)\,\sum_{n\in\Z}\ e^{\ii\,n(\gamma-\frac{\varphi}{2})}\\
 &= \frac{1}{(2\pi)^2}\int_{\mathbb{S}^{1}}\ud\gamma\,\int_{\mathbb{S}^{1}} d\varphi\,e^{\ii\,m\,\varphi}\overline{\psi(\gamma+\theta)}
\psi(\gamma-\varphi+\theta)\, {(2\pi)\,\delta\left(\gamma-\frac{\varphi}{2}\right)} \\
 &=  \frac{1}{2\pi}\int_{\mathbb{S}^{1}} \ud\varphi\,e^{\ii\,m\,\varphi}\overline{\psi\left(\theta+\frac{\varphi}{2}\right)}
\psi\left(\theta-\frac{\varphi}{2}\right)\,.
\end{align*}
\eprf
Given a state $\psi$, one can also show that its Wigner distribution can be retrieved  through {the symplectic Fourier transform}  of  its reproducing kernel, that is:
\begin{align*}
&\frac{1}{(2\pi)^2}\sum_{m\in\Z}\int_{\mathbb{S}^{1}}\ud\gamma\,e^{\ii\,(m^{\prime}\theta-m\theta^{\prime})}\lg\phi_{(m^{\prime},\theta^{\prime})}\vert\psi\rg\\\nonumber
&=\frac{1}{(2\pi)^2}\sum_{m^{\prime}\in\Z}\int_{\mathbb{S}^{1}}\ud\theta^{\prime}\,e^{\ii\,(m^{\prime}\theta-m\theta^{\prime})}\int_{\mathbb{S}^{1}}\ud\gamma\, e^{\ii\frac{\,m^{\prime}\theta^{\prime}}{2}}e^{-\ii\,m^{\prime}\gamma}\overline{\psi(\gamma-\theta^{\prime})}\psi(\gamma)\\\nonumber
&=\frac{1}{(2\pi)^2}\int_{\mathbb{S}^{1}}\ud\theta^{\prime}\,e^{-\ii\,m\theta^{\prime}}\int_{\mathbb{S}^{1}}\ud\gamma\, \overline{\psi(\gamma-\theta^{\prime})}\psi(\gamma)2\pi\delta\left(\theta+\frac{\theta^{\prime}}{2}-\gamma\right)\\\nonumber
&=\frac{1}{2\pi}\int_{\mathbb{S}^{1}}\ud\theta^{\prime}\,e^{-\ii\,m\theta^{\prime}}\, \overline{\psi\left(\theta-\frac{\theta^{\prime}}{2}\right)}\psi\left(\theta+\frac{\theta^{\prime}}{2}\right)\,.
\end{align*}
{\begin{remark}
We observe that the Wigner function for $\psi$, which is  given in Eq.\;\eqref{wignerpsi},  is the $m$-th Fourier coefficient of the $2\pi$-periodic complex valued function $\overline{\psi\left(\theta-\frac{\cdot}{2}\right)}\psi\left(\theta+\frac{\cdot}{2}\right)$. Due to Eq.\;\eqref{residMotr} and the normalisation $\Vert\psi\Vert^2=1$ we have 
\begin{equation}
\label{wignerquasi}
\frac{1}{2\pi}\sum_{\Z}\int_{\mathbb{S}^1}\ud\theta\, W_\psi(m,\theta)=1\, .
\end{equation}  
Hence, like for the standard  phase space $\R\times\R$, it is a normalised, not necessarily non-negative  distribution on the discrete cylinder \cite{mukunda1}. However, while the Wigner function for the former case is non-negative for Gaussian states, and more generally for standard coherent states, it is not true in the case of the discrete cylinder, as is well asserted by the recent study  \cite{Kowalski2} on the  circular CS, the Gaussian CS and the Gaussian–Fourier on the circle. As a matter of fact it was proved in \cite{Rigas10} that the Wigner function of a pure state $|\psi \rg$ is non-negative if and only if $|\psi\rg$ is an eigenstate of the angular momentum operator.
\end{remark}}  

{In this case of unit weight} $\varpi_{\sfP}=1$, we trivially have for $\Omega(\gamma)= \varpi_{\sfP}(0,\gamma)=1$: $\Omega^{\prime}(\gamma)=\Omega^{\prime\prime}(\gamma)=0$. 
Hence we get for the quantizations of elementary functions, 
\begin{itemize}
\item $g(m)=m$, $A^{\varpi_\sfP}_{m}=L$,
\item $g(m)=m^2$, $A^{\varpi_\sfP}_{m^2}=L^2$, 
\item $h(\theta)$, 
\begin{equation}
(A^{\varpi_{\sfP}}_h\psi)(\gamma)= \left(\sum_{m\in\Z} \hat{h}(m)\, \frac{e^{\ii\,m\gamma}}{\sqrt{2\pi}}\right)\psi(\gamma)= h(\gamma)\psi(\gamma)\,.
\end{equation}
which means that:
\begin{equation}
(A^{\varpi_\sfP}_{\sin\theta}\psi)(\gamma)= \sin\gamma \,\psi(\gamma)\, , \quad  (A^{\sfP}_{\cos\theta}\psi)(\gamma)= \cos\gamma \,\psi(\gamma)\, . 
\end{equation}
\end{itemize}
In this case, we can consider the position operator through the multiplication by the smooth Fourier exponential $e^{\ii\gamma}$. 

Alternatively we can  quantize the periodized  angle function  $(\mathrm{Per}\,\mathsf{A})(\theta)$. 
One finds the multiplication operator defined by the same discontinuous $(\mathrm{Per}\,\mathsf{A})(\gamma)$. There is no regularisation. On the contrary  the choice of a smooth weight function allows to obtain a multiplication by smooth regularisation of this angle function. 

Finally, for the angular parity weight, the expression in Eq.\;\eqref{distvap} reduces to $$\mathrm{Tr}\left(\sfM^{\varpi_\sfP}(m,\theta )\sfM^{\varpi_\sfP}\right)=\delta_{m 0}\delta(\theta)\,.$$ 
In this  case  the semi-classical portrait of $A^{\varpi_\sfP}_f$ is $f$, $\check f(m,\theta)= f(m,\theta)$.

\section{Conclusion}
\label{conclu}

In this paper, we have established a covariant  integral quantization {for  systems moving on the circle with integral momentum, i.e, for systems whose  phase space is {the discrete cylinder} $\Z\times\,\mathbb{S}^1$}. 
The symmetry  group of this phase space is the discrete \& compact  version of the Weyl-Heisenberg group, namely  the central extension of the abelian group $\Z\times\,\mathrm{SO}(2)$,  and the phase space can be viewed as the {right} coset of the group with its center. The existence of a non-trivial unitary irreducible representation  of this group on the phase space, as acting on $L^2(\mathbb{S}^1)$, allowed us to derive interesting results. First, we have established the concomitant resolution  of the identity and subsequent properties such as the Gabor transform on the circle and its inversion, the reproducing kernel and the fact that any square integrable  function on the circle is  fiducial vector. {Moreover, picking a weight function on the discrete cylinder with suitable properties allows to build a bounded self-adjoint operator on $L^2(\mathbb{S}^1)$. The Weyl transported versions of this operator yield a resolution of the identity and the subsequent covariant quantization of functions or distributions defined on the discrete cylinder. }

There are noticeable results related to the quantization of a point $(m,\theta)$ in the phase space according to two standard choices of the weight. 

\begin{itemize}
\item With the parity weight,  the quantization of the momentum is the expected angular momentum operator $L$. 
\begin{equation}
m\mapsto\hat{m}= L\, , \quad L\psi(\gamma)= -\ii\frac{\partial}{\partial\gamma}\psi(\gamma)\,.
\end{equation}
while the quantization of the angle yields the multiplication operator by the angle.
\begin{equation}
\theta\mapsto\hat{\theta}\, , \quad \hat{\theta}\psi(\gamma)= \gamma \psi(\gamma)\,. 
\end{equation}
 This is of course not acceptable. Alternatively one can quantize the periodized  angle function  $(\mathrm{Per}\,\mathsf{A})(\theta)$. One finds the multiplication operator defined by the same discontinuous $(\mathrm{Per}\,\mathsf{A})(\gamma)$. There is no regularisation.
\item With the coherent state weight one obtains the quantization  of the momentum as the usual $L$ plus an additional term, i.e.,  a kind of covariant derivative on the circle whose topology is now taken into account,
\begin{equation}
m\mapsto\hat{m}= L+\lg\,m\rg_{\vert\psi\vert^2}\, , 
\end{equation}
whereas the quantization of the periodized function theta leads to its smooth regularisation,  
\begin{equation}
(\mathrm{Per}\,\mathsf{A})(\theta)\mapsto\left[ \sum_{m\in\Z} \hat{h}(m)\,\lg e^{-\ii m\gamma}\rg_{\vert\psi\vert^2}\,e^{\ii\,m\gamma}\right]\phi(\gamma)\,. 
\end{equation}

\end{itemize}

%
In a forthcoming work, we will extend the results of the present paper in three directions.
\begin{itemize}
\item Analyzing circular data (see for instance \cite{landler,aquino,Asorey}).
We expect that the formalism we have developed above will be useful for  circular data or circular statistics.  Some of the fiducial vectors we have considered here are  probability densities in these areas, namely  the uniform distribution, the shifted gaussian, and the Von Mises, Fe\'jer and Poisson.  
\item Quantum systems for which the configuration space is SO$(3)$: the group of rotations in three dimensions \cite{risbo}. 
{In this case \cite{mukunda2}, although  the corresponding phase space is not a coset arising from a group, the Weyl formalism still applies}.
\item Stellar representation.
{In a forthcoming  work \cite{fabre22}} we intend to link this formalism for both SO(2) and SO(3) to the so-called stellar representations \cite{majorana32,chabaud20} .
\end{itemize}

\subsection*{acknowledgments}

J.-P. Gazeau thanks  the ICTP Trieste for financial support and  hospitality.

\appendix
\section[\appendixname~\thesection]{Some fiducial vectors \& reproducing kernels}
\label{table}

For simplicity in the table below we denote $A(m,m^{\prime},\theta,\theta^{\prime})$ by $A$:

\begin{center}
    \begin{tabular}{ | l | l | l | l | l  | l |   l | p{10 cm} |}
  \hline

 &Fiducial vector& Reproducing kernel	                                                                                     \\ \hline

{General} & $\phi(\gamma)$ & $\lg\phi_{(m,\theta)}\vert\phi_{(m^{\prime},\theta^{\prime})}\rg$ 	   \\ \hline

{Constant} & $\frac{1}{\sqrt{2\pi}} $   & $\tiny{\delta_{m\,m^{\prime}}e^{\ii\,m\,(\theta-\theta^{\prime})}}$ 					       \\ \hline

{Basis} & $\frac{1}{\sqrt{2\pi}}\times\,e^{\ii\,n\gamma}$  & $A\,\delta_{m\,m^{\prime}}$ 			       \\ \hline

{Shifted gaussian} & ${(2\pi)}^{-\frac{1}{2}}\theta\left(\frac{\gamma}{2\pi},\frac{\sigma}{\sqrt{2\pi}}\right)$   & $\tiny{4\pi^2A\,e^{-\frac{(m-m^{\prime})^2}{2}}\sum_{n\in\Z}\,e^{-n^2}e^{\ii\,n\,(\ii(m-m^{\prime})+(\theta-\theta^{\prime}))}}$  				        \\ \hline
{Dirichlet} & $\frac{\sin(n+\frac{1}{2})\varphi}{\sin{\frac{1}{2}}\varphi}$   & $\frac{A}{2\pi(n+1)}\,\frac{\sin((n+\frac{1}{2})(\theta-\theta^{\prime}))}{\sin(\frac{\theta-\theta^{\prime}}{2})}
$ 				          \\ \hline
{Fejer}& $\frac{1}{n+1} \left[{\frac{\sin((\frac{n+1}{2})\gamma)}{\sin{\frac{1}{2}\gamma})}}\right]^2\,$ &$A\,\sum_{k=-n}^{k=n}\left(1-\frac{\vert\,k\vert}{n+1}\right)\left(1-\frac{\vert\,k-(m-m^{\prime})\vert}{n+1}\right)\,e^{\ii\,k(\theta-\theta^{\prime})}$ 				          \\ \hline
{Von  Mises} & $\frac{e^{\lambda\cos(\gamma)}}{\sqrt{2\pi\,I_{0}(2\lambda)}}$  &$\frac{A}{2\pi\,I_{0}(2\lambda)\!\!}\,I_{m-m^{\prime}}\left(2\lambda\,\cos\left(\frac{\theta-\theta^{\prime}}{2}\right)\right)$ 				     \\    \hline
 \hline
    \end{tabular}
\end{center}


\begin{thebibliography}{999}


\bibitem{Carruthers}  Carruthers, P.  and  Nieto, M.~M. Phase and angle variables in quantum mechanics, \emph{Rev. Mod. Phys.} \textbf{1968}, \emph{40}, 411-440.

\bibitem{leblond}  L\'evy-Leblond, J.-M. Who is afraid of nonhermitian operators? A quantum description of angle  and phase,  \emph{Ann. Phys. (NY)} \textbf{1976}, \emph{101}, 319-341. 

\bibitem{mukunda1} Mukunda, N.  Wigner distribution for angle coordinates in quantum mechanics, \emph{Am. J.  Phys.} \textbf{1979}, \emph{47}, 182-187.

\bibitem{Floreanini} Floreanini, R.;   Peracci, R.; Sezgin, E.  Quantum mechanics on the circle and $W_{1+\infty}$, \emph{Phys. Lett. B} \textbf{1991}, \emph{271},  372-376.

\bibitem{hall} Hall, B.~C.  and  Mitchell, J.~J. Coherent states on spheres, \emph{J. Math. Phys.} \textbf{2002}, \emph{43}, 1211-1236.

\bibitem{Scardicchio}  Scardicchio, A. Classical and quantum dynamics of a particle constrained on a circle, \emph{Phys. Lett. A} \textbf{2002}, \emph{300}, 7-17.

\bibitem{Gour} Gour, G. The quantum phase problem: steps toward a resolution, \emph{Found. Phys.} \textbf{2002}, \emph{32}, 907-926.

\bibitem{Zhang}  Zhang, S. and Vourdas, A.  Phase space methods for particles on a circle, \emph{J. Math. Phys.} \textbf{2003}, \emph{44}, 5084-5094.

\bibitem{Kowalski1}  Kowalski, K.;  Rembielinski, J.;  Papaloucas, L.~C. Coherent states for a quantum particle on a circle, \emph{J. Phys. A: Math. Gen.} \textbf{1996} \emph{29}, 4149-4167.

\bibitem{debievre}  De Bi\`evre, S. Coherent states over symplectic homogeneous spaces,  \emph{J. Math. Phys.} \textbf{1989}, \emph{30}, 1401-1407.

\bibitem{aremua} Aremua, I.; Gazeau, J.-P. ;  Hounkonnou, M.~N. Action-angle coherent states 
for quantum systems with cylindric phase space, \emph{J. Phys. A: Math. Theor.} \textbf{2012}, \emph{45}, 335302-01-16.

\bibitem{Rigasa} Rigas, I.;   Sanchez-Soto, L.~L.;  Klimov, A.~B.;  \v{R}eh\'a\v{c}ek, J.;  Hradil, Z. Orbital angular momentum in phase space, \emph{Ann.  Phys.} \textbf{2011}, \emph{326},  426-439.

\bibitem{chadzitaskos}  Chadzitaskos, G.;  Luft, P.;   Tolar, J. Quantizations on the circle and coherent states, \emph{J. Phys. A: Math. Theor.} \textbf{2012}, \emph{45} 244027-1-15.

\bibitem{Przanowski}  Przanowski, M.;  Brzykcy, P.;   Tosiek, J. From the Weyl quantization of a particle on
the circle to number-phase Wigner functions, \emph{Ann. Phys.} \textbf{2014}, \emph{351},  919-934.

\bibitem{kastrup} Kastrup, H.~A.  Wigner functions for the pair angle and orbital angular momentum, \emph{Phys. Rev. A} \textbf{2016}, \emph{84},062113-01-14.

\bibitem{Fresneda}  Fresneda, R.;   Gazeau, J.-P.;  Noguera, D. Quantum localisation on the circle, \emph{J. Math. Phys.} \textbf{2018}, \emph{59}, 052105-01-19.

\bibitem{Kowalski2}  Kowalski, K. and   Lawniczak, K.  Wigner functions and coherent states for
the quantum mechanics on a circle, \emph{J. Phys. A: Math. Theor.} \textbf{2021}, \emph{54}, 275302-01-23.

\bibitem{perel86}  Perelomov, A.M. \textit{Generalized Coherent States and Their Applications}, Springer, Berlin, 1986.

\bibitem{Gonzales}  Gonzalez, J.~A. and  del Olmo M.~A., Coherent States on the Circle, \emph{J. Phys. A: Math. Gen.} \textbf{1998}, \emph{31}, 8841-8857.

\bibitem{Mista}  Mista Jr., L.;  de Guise, H.; \v{R}eh\'a\v{c}ek, J.;  Hradil, Z. Angle and angular momentum: Uncertainty relations, simultaneous measurement, and phase-space representation, 
\emph{Phys. Rev. A.} \textbf{2022}, \emph{106}, 022204-1-12; arXiv:2201.02231 [quant-ph]


\bibitem{bergaz14} {Bergeron, H. and   Gazeau, J.-P., 
Integral quantizations with two basic examples,  \emph{Annals of Physics (NY)} \textbf{(2014)},  \emph{344},  43-68.}


\bibitem{Richard_AU}  Laugesen, R. S. \emph{Harmonic Analysis Lecture Notes}; 	arXiv:0903.3845 [math.CA]

\bibitem{gazmur16} Gazeau, J.-P.  and Murenzi, R.  Covariant Affine Integral Quantization(s),  
\emph{J. Math. Phys.} \textbf{2016}, \emph{57}, 052102-1-22;  arXiv:1512.08274 [quant-ph]

\bibitem{gazkoimur20} Gazeau, J.-P.;  Koide, T.;  Murenzi, R. 2-D Covariant Affine Integral Quantization(s), 
\emph{Adv. Oper. Theory} \textbf{2020}, \emph{5}, 901-935.   

\bibitem{gazkoimur22} Gazeau, J.-P.;  Koide, T.;  Murenzi, R.,  Correction to: 2-D Covariant Affine Integral Quantization(s), 
\emph{Adv. Oper. Theory} \textbf{2020}, \emph{7}, 22.

\bibitem{marsden74}  {Marsden, J. and Weinstein, A. Reduction of Symplectic manifolds with symmetry, \emph{Rep. Math. Phys.} \textbf{1974}, \emph{5}, 121-129}

\bibitem{grossmann} Grossmann, A.  Parity Operator and Quantization of $\delta$-Functions, \emph{Comm. Math. Phys.} \textbf{1976}, \emph{48}, 191-194. 

\bibitem{Rigas10} {Rigas, I.;   Sanchez-Soto, L.~L.;  Klimov, A.~B.;  \v{R}eh\'a\v{c}ek, J.;  Hradil, Z. Non-negative Wigner functions for orbital angular momentum states, \emph{Phys. Rev A} \textbf{2010}, \emph{81}, 012101-1-4.}

\bibitem{landler}  Landler, L.;  Ruxton, G.~D.;   Malkemper, E.~P. Circular data in biology: advice for effectively implementing statistical procedures, \emph{Behavioral Ecology and Sociobiology} \textbf{2018}, \emph{72}, 128-01-10.

\bibitem{aquino}  Aquino, M.~B.~L.;  Guimaraes, J.~P.~F.;  Fontes, A.~I.~R.;  Linhares, L.~L.~S.;  Rego, J.~B.~A.;
 De M. Martins, A. Circular Correntropy: Definition and Application
in Impulsive Noise Environments, \emph{IEEE Access} \textbf{2022}, \emph{10}, 58777-58786.

\bibitem{Asorey}  Asorey, M.;  Facchi, P.;  Man'ko, V.~I.;  Marmo, G.;  Pascazio, S.;  Sudarshan, E.~G.~C.
Radon transform on the cylinder and tomography of a particle on the circle, \emph{Phys. Rev. A} \textbf{2007}, \emph{76},  012117-01-09.


\bibitem{risbo}   Risbo, T. Fourier transform summation of Legendre series and $D$-functions, \emph{J. Geodesy} \textbf{1996}, \textbf{70}, 383-396.

\bibitem{mukunda2}  Mukunda, A.;  Chaturvedi, S.;   Simon, R.  Wigner distributions and quantum mechanics on Lie groups: the
case of the regular representation, \emph{J. Math. Phys.} \textbf{2004}, \emph{45}, 114-148.

\bibitem{fabre22}  {Fabre, N.; Gazeau, J.-P.; Murenzi, R.;  Sanchez Soto, L.~L., \textit{in progress}.}

\bibitem{majorana32} {Majorana, E. Teoria relativistica di particelle con momento intrinseco arbitrario,   \emph{Il Nuovo Cimento} \textbf{1932}, \emph{9}, 335-344.}

\bibitem{chabaud20}  {Chabaud, U;  Markham, D;  Grosshans, F.  Stellar Representation of Non-Gaussian Quantum States, 
\emph{Phys. Rev. Lett.} \textbf{2020}, \emph{124}, 063605-1-6.} 

\end{thebibliography}
\end{document}